\documentclass[prb,twocolumn,showpacs,superscriptaddress,footinbib]{revtex4}


\newcommand{\beq}{\begin{equation}}
\newcommand{\eeq}{\end{equation}}
\newcommand{\beqn}{\begin{eqnarray}}
\newcommand{\eeqn}{\end{eqnarray}}

\newcommand{\slp}{\raise.15ex\hbox{$/$}\kern-.57em\hbox{$ \partial $}}
\newcommand{\lnA}{\raise.15ex\hbox{$/$}\kern-.57em\hbox{$A$}}

\usepackage{graphicx}
\usepackage{epsfig}
\usepackage{bm}

\begin{document}
\title{Conformal properties of 1D quantum systems with long-range interactions}

\author{Carlos M.\ Na\'on}
\author{Mariano J.\ Salvay}
\author{Marta L. Trobo}
\affiliation{Departamento de F\'\i sica, Facultad de Ciencias
Exactas, Universidad Nacional de La Plata, CC 67, 1900 La Plata,
Argentina} \affiliation{Instituto de F\'\i sica La Plata, Consejo
Nacional de Investigaciones Cient\'\i ficas y T\'ecnicas,
Argentina}

\date{August, 18 2005}

\begin{abstract}
We investigate conformal properties of a one-dimensional quantum
system with a long-range, Coulomb-like potential of the form
$\frac{1}{|x|^{\sigma}}$, with $\sigma >0$. We compute the
conformal anomaly $c$ as function of $\sigma$. We obtain $c=0$ for
$\sigma <1$ (Wigner crystal) and $c=1$ for $\sigma >1$ (Luttinger
liquid). By studying the scaling regime of the system when it is
deformed by the inclusion of a term $t\,\rho(x)$, (with $\rho(x)$
a scaling operator), we show that, for $\sigma >1$, the
correlation length $\xi$ gets an additional interaction dependent
factor. This leads to a necessary redefinition of $\xi$ in order
to avoid an unphysical dependence of the central charge on the
coupling constant.
\end{abstract}

\maketitle
\section{Introduction}
The principle of conformal invariance (CI) at a critical point is
one of the most powerful tools for our understanding of
one-dimensional (1D) quantum systems \cite{reviews}. This
invariance constrains the possible behavior of a given critical
theory, allowing a characterization of universality classes in
terms of a dimensionless parameter c (the conformal anomaly) which
is the central charge of the quantum extension of the
(infinite-dimensional) conformal algebra. The central charge gives
a measure of the sensitivity of a critical theory when one changes
the space geometry. It has also been related to the
low-temperature specific heat in quantum chains
\cite{Affleck-Cardy}. As a striking achievement, in the context of
classical two-dimensional statistical mechanics models, conformal
invariance led to the derivation of all critical exponents
corresponding to theories with $c < 1$ \cite{FQS}. When the models
posses an additional continuous symmetry a classification of
critical models is also possible for $c \geq 1$ \cite{Cardy3}.

CI is obtained by assuming translational, rotational and scale
invariance, together with short-range interactions. If any of
these conditions are not verified its validity can not be assured.
However, the concepts of CI have been extended in several
directions. Cardy, has derived a relationship between the second
moment of the correlation function of a scaling operator away from
criticality (in the scaling region) and the central charge and the
scaling dimension corresponding to the critical theory
\cite{Cardy4}. CI has been also applied, as a principle, to 1D
quantum systems with inverse-square long-range interactions
\cite{Sutherland1} \,\ \cite{Haldane1}. In particular, the
continuum limit of the $S=1/2$ Heisenberg chain with
inverse-square exchange was associated to a $c=1$ $SU(2)$
Kac-Moody theory \cite{Haldane2} and Sutherland's model with
interaction $g/r^2$ \,\, \cite{Sutherland1} was analyzed through
finite size scaling \cite{Kawakami} \,\, \cite{Sutherland2}.
Finite size corrections to the ground state led to a g-dependent
central charge, whereas the low temperature expansion of the free
energy yielded $c=1$, which is consistent with the scaling
dimensions obtained for density operators and momentum
distributions. The reason for this discrepancy has not been fully
explained, although it was suggested that it might be related to
the finite-size approximation to the long-range potential
\cite{Kawakami} and to the possible differences between time and
space correlations in long-ranged models \cite{Sutherland2}.

On the other hand, in the context of low-dimensional condensed
matter systems, the interest in long-range interactions is not a
purely academic one. For instance, in quasi-one-dimensional
systems one usually considers contact electron-electron
interactions assuming that the presence of adjacent chains
effectively screens the basic Coulomb interaction. However, as the
dimensionality of the system decreases charge screening effects
become less important and consequently the long-range interaction
plays an observable role \cite{longrange}. The effect of power-law long-range interactions is also of interest for the study of antiferromagnetic spin-chains \cite{spin-chains}, in which possible experimental realizations might be found, for instance, in systems with Ruderman-Kittel-Kasuya-Yosida interactions \cite{RKKY}.

The purpose of this paper is to explore the interplay between the
principle of conformal invariance and the effect of long-range,
distance-dependent interactions. By performing analytical
computations of the central charge c in a continuous theory, we
shed light on the effect of anisotropy on the conformal properties
of a 1D quantum system. In particular, in connection to the above
mentioned discrepancy, we will explicitly show how the asymmetry
between long-time and long-space behaviors of the correlations
must be treated in order to obtain sensible results for the
conformal anomaly.

Most of the previous calculations of c for long-ranged systems have been done on the
basis of a discrete model and for specific values of $\sigma$. The
exception is the important paper by Mironov and Zabrodin
\cite{Mironov-Zabrodin}, who were able to determine scaling
dimensions and correlation functions for a general class of
potentials. However, in their formulation it was not possible to
specify the conditions to be satisfied by the potentials in order
to assure the absence of a gap. Moreover, the value of c was not
computed but inferred from the form of some scaling dimensions.

\section{The model and the long-range potential}

We will study a bosonic non local (1+1)-dimensional quantum field
theory with Lagrangian density

\beq {\cal L}_0
=\frac{1}{2}(\partial_{\mu}\phi(x))^2\,+\,\frac{1}{2} \int
d^2y\,\partial_{1}\phi(x)\,V(x-y)\,\partial_{1}\phi(y). \label{6}
\eeq

This model is the bosonized version of a
Thirring-Tomonaga-Luttinger fermionic model containing only
forward-scattering interactions \cite{Li-Naon}. It describes
charge-density oscillations with sound velocity
$v=\sqrt{1+V(p_1)}$, where $V(p_1)$ is the Fourier transform of the distance-dependent potential. Although the model under consideration is not Lorentz-covariant, we shall use, for convenience, a notation which is reminiscent of relativistic problems ($d^2x=dx_0 dx_1$, $d^2p=dp_0 dp_1$).

We will consider an instantaneous, power-law decaying potential of
the form

\beq V(x)=g\,\delta(x_0)\,(x_1^2 + \alpha^2)^{\frac{-\sigma}{2}}
\label{1}, \eeq where $\sigma$ is an arbitrary positive real
number and $\alpha$ is a regulator parameter that can be
associated to the lattice spacing.

Since we shall deal with all positive values of $\sigma$ we will
need the Fourier transformed potential which depends on $\sigma$
as follows:

\beq V(p_{1})= \frac{2g
\sqrt{\pi}(2\alpha)^{(1-\sigma)/2}}{\Gamma(\sigma/2)}
|p_1|^{(1-\sigma)/2} K_{(1-\sigma)/2}(\alpha |p_1|), \label{2}\eeq
for $\sigma > 1$. Here $K_{(1-\sigma)/2}(\alpha \mid p_1 \mid)$ is
a modified Bessel function. Note that, in this case, the potential
is renormalized. This renormalization is associated to the divergence of the integrand in the Fourier integral, for $x_1=0$. This divergence is not present when $\sigma < 1$.
Defining the renormalized coupling constant
$g_R=(\alpha)^{1-\sigma}g$ and taking into account the behaviour
of the Bessel function for small arguments we get \beq V(p_{1})=
f_> (g_R, \sigma),\label{3} \eeq with $f_> (g_R,
\sigma)=\frac{g_R\, \sqrt{\pi}\,\Gamma((\sigma
-1)/2)}{\Gamma(\sigma/2)}$.

For $\sigma < 1$ the Fourier-transformed potential reads \beq
V(p_{1})=f_<(g, \sigma)\,|p_1|^{(\sigma -1)} \label{4},\eeq with
$f_<(g, \sigma)=\frac{g
\,\sqrt{\pi}\,(2)^{1-\sigma}\,\Gamma((1-\sigma)/2)}{\Gamma(\sigma/2)}
$.

As we shall see, the Coulomb potential corresponding to $\sigma=1$
is a special case that can be considered as the frontier between
two regions of qualitatively different conformal behaviors. This
case was investigated, from another point of view, by Schulz, in
his work on Wigner crystal-like formation in a Luttinger system
\cite{Schulz} (See also ref. 17 for an explicit computation of the electronic single-particle function). For $\sigma=1$ one has \beq V(p_{1})=
2\,g\,K_0(\alpha |p_1|) \approx -2\,g\,\ln(\alpha
|p_1|)\label{5}.\eeq

\section{Computation of the conformal anomaly}
  Exactly as it was done in Refs. 9 and 10 with
the Sutherland's model, for the special case $\sigma=2$, in this
paper we shall employ the model given by (\ref{6}) (for arbitrary
real and positive $\sigma$) as a testing ground to examine the
consequences of imposing CI in the presence of anisotropic
interactions. Our idea is to compute the central charge following
two different routes. Firstly, we will compute the low-temperature
behaviour of the specific heat, from which we can read the value
of c, according to the results of Refs. 2. Secondly we will
perturb the critical Lagrangian (\ref{6}) by adding a simple
scaling operator $\rho(x)$ that takes the system away from
criticality. Then, by considering the second moment of the
two-point correlations of $\rho(x)$ in the scaling regime, we can
obtain the value of the conformal anomaly by using the result
derived by Cardy \cite{Cardy4}:

\begin{equation}
\int d^2x\,|x|^2 \langle\rho(x)\rho(0)\rangle_t = \frac{c}{3\, \pi
\,t^2\,(2 - \Delta_\rho)^2} \label{7},
\end{equation}
where $\Delta_\rho$ is the scaling dimension of $\rho$ and $t$ is
the coupling constant of the added term that spoils CI
($t\propto\xi^{-1}\propto(T-T_c)$). Although somewhat indirect,
this method is interesting because it involves both large spatial
and temporal dependence of the correlations. This fact will enable
us to unravel some subtleties involved in the application of CI
concepts to anisotropic models. Let us point out that, eventhough the above equation is valid for any scaling operator, $\rho(x)$ is the density operator throughout this work.

\subsection{First method: free-energy calculation}
In order to compute the low temperature behavior of the specific
heat we study (\ref{6}) in the imaginary-time formalism
\cite{finite-temperature}. As usual, we build the partition
function ${\cal Z}$ as a functional integral extended over the
paths with periodicity conditions in the Euclidean time variable
$x_0$: $\phi(x_0 + \beta, x_1) = \phi( x_0, x_1)$, where
$\beta=\frac{1}{T}, k_B =1$. After some standard algebraic steps
(see Ref. 15 for a detailed similar computation in a fermionic
model) the Helmholtz free energy ${\cal F}=- \frac{1}{\beta} \ln
{\cal Z}$ can be expressed as
\begin{equation}
{\cal F} = {\cal F}_0  +\frac{1}{\beta}
\int_{-\infty}^{\infty}\frac{dp_1}{2 \pi}\, \ln ( 1 - e^{- \beta
\, |p_1| \,v(p_1)}), \label{8}
\end{equation}
where ${\cal F}_0$ is the zero-point, vacuum free energy.
Now we have to evaluate the integral and read the value of $c$.
This is easily done for $\sigma>1$, since
$v=\sqrt{1+f_>(g_R,\sigma)}=v(g_R,\sigma)$ is independent of the
momentum. The result is
\begin{equation}
{\cal F} = {\cal F}_0  -\frac{\pi}{6\,\beta^2\,v(g_R,\sigma)}
\label{9},
\end{equation}
which means that $c=1$$\,$ \cite{Affleck-Cardy}. For $\sigma<1$
the computation is a little bit more subtle because the sound
velocity depends on $p_1$. However, taking into account that one
is interested in the large-$\beta$ behaviour of ${\cal F}$, the
integral can be approximated, yielding a result proportional to
$\beta^{-\frac{\sigma +3}{\sigma+1}}$. For $\sigma<1$ the exponent
$\frac{\sigma +3}{\sigma+1}>2$, from which one concludes that
$c=0$. Conformal field theories with $c=0$ have been previously
found in the study of second order phase transitions in the
presence of quenched disorder \cite{Cardy5}. Some of their
interesting properties have been analyzed quite recently
\cite{Gurarie}.

\subsection{Second method: deformation of a conformal field theory}

Let us now undertake the computation of $c$ following the second
method described above. We choose as perturbing operator
$\rho(x)=\partial_{1}\phi(x)$. In order to determine the scaling
dimension we compute the critical two-point correlation. A
straightforward manipulation yields

\beq \langle\rho(x)\,\rho(0)\rangle_0 =
\frac{1}{(2\pi)^2}\,\int\,d^2p\,\frac{p_1^2\,e^{ip.x}}{p_0^2 +
p_1^2\,v^2(p_1)} .\label{10} \eeq

As a first step we evaluate the scaling dimensions by considering
equal-time correlations. Performing a long-distance approximation
in the above integral we get \beq
\langle\rho(x_1)\,\rho(0)\rangle_0 =
\frac{1}{4\,\pi\,v}\,\frac{1}{|x_1|^2} \label{11}, \eeq for
$\sigma>1$, from which we obtain $\Delta_\rho = 1$. Note that in
this case $v=\sqrt{1+f_>(g_R,\sigma)}=v(g_R,\sigma)$.

The analogous calculation for $\sigma<1$ gives a lengthy
expression in terms of a confluent hypergeometric function.
Analyzing the asymptotic long distance limit we obtain \beq
\langle\rho(x_1)\,\rho(0)\rangle_0 \sim
\,\frac{1}{\pi\,(5-\sigma)\,\sqrt{f_<(g,\sigma)}\,|x_1|^{2(1+\frac{1-\sigma}{4})}}
\label{12}, \eeq from which we get $\Delta_\rho =
1+\frac{1-\sigma}{4}$. Note that both results, for $\sigma>1$ and
$\sigma<1$ coincide for $\sigma=1$, which suggests that
$\Delta_\rho$ is continuous for all $\sigma
> 0$. We have explicitly verified the correctness of this assertion: \beq
\langle\rho(x)\,\rho(0)\rangle_0 =
\frac{e^{1/g}\,\sqrt{\pi/2g}}{\alpha^2}\,Erfc[\sqrt{\frac{1}{g}-2\ln{\frac{\alpha}{|x_1|}}}],\label{13}
\eeq where Erfc stands for the complementary error function. In
the long-distance limit this expression behaves as $|x_1|^{-2}$,
i.e. $\Delta_\rho =1$ for $\sigma =1$. We will use this result
later.

At this point we start considering the modified, non critical
theory described by

\beq {\cal L} ={\cal L}_0\,+\,t\,\rho(x), \label{14} \eeq and try
to compute the left hand side of (\ref{7}). This computation
involves correlations slightly away from criticality, in the
scaling regime in which both $|x_1|$ and the correlation length
$\xi$ are very large, but with $|x_1| <<\xi$. Besides, both
spatial and temporal dependence of the correlations are needed. In
fact, regarding the possible application of CI to an anisotropic
model such as the one considered here, this is a crucial point.
Indeed, it is very instructive to compute again
$\langle\rho(x)\,\rho(0)\rangle_0$, not in the equal-time case but
for $x_0, x_1 \neq 0$. This is quite easily done for $\sigma>1$.
The result is

\beq \langle\rho(x)\,\rho(0)\rangle_0 = \frac{e^{-|x_0|\,v/|x_1|}
(e^{|x_0|\,v/|x_1|}\,-1\,-\frac{|x_0|\,v}{|x_1|})}{2\,\pi\,v^3\,|x_0|^2}
\label{15}. \eeq Performing an expansion of the numerator around
$x_0=0$ we reobtain our previous result for the equal-time
correlator, equation (\ref{11}). The important observation here is
that although one can define a scaling dimension which is uniquely
determined for large distances and long times, the precise
functional form of the correlation {\it depends on the direction
in which large scales are observed}. The interesting point is that
in contrast to previous calculations, our  procedure is able to
take into account this expected anisotropy. Since the integrand in
the left hand side of (\ref{7}) involves the general (not
equal-time) behavior of the correlator, our result for c will
reflect the contributions coming from different directions of
space-time. In view of these remarks we now compute the left hand
side of (\ref{7}). In our calculation the correlation length is
associated to spatial fluctuations only. This parameter is used as
an infrared cutoff for the spatial integral. For $\sigma<1$ we get
\beq \int d^2x\,|x|^2 \langle\rho(x)\rho(0)\rangle_t =
C_1\,\xi^{\sigma} +C_2\,\xi^{\sigma+1}+C_3\,\xi^{2\,\sigma}
\label{16}, \eeq where $C_1$, $C_2$ and $C_3$ are constants
depending on $g$ and $\sigma$. Since $t \propto \xi^{-1}$ one sees
that $c\rightarrow 0$, in agreement with the result obtained by
computing the free energy (\ref{8}). It is more illuminating the
case $\sigma>1$ which gives \beq c = (\frac{1}{v^2} +
\frac{6}{v^4})\,\xi^{2}\,t^2 \label{17}. \eeq Since we know from
(\ref{9}) that $c=1$, we obtain an expression for the correlation
length in terms of $t$ and $v$. The extra dependence of $\xi$ on
$v$ is due to the anisotropy of the system, although the precise
form of this dependence is a consequence of our regularization
prescription. Please note that if we assume that $\xi$ is a
function of $t$ only, the above equation predicts a conformal
anomaly that depends on the long-range potential through the sound
velocity, in analogy with the results previously obtained in Refs.
9 and 10 (for a different model and for $\sigma =2$). This
situation reflects the fact that the correlation length is defined
up to a multiplicative non-universal constant.

For the particular value $\sigma =1$ the above procedures are not
easy to implement due to the peculiar form of the potential
(\ref{5}). However, based on the results (\ref{11}) and (\ref{13})
it is straightforward to conclude that $c=1$ for $\sigma =1$ also.

\section{Summary and conclusions}

In summary, we have presented the first explicit and analytical
computation of the conformal anomaly in a system with long-range
Coulomb-like interactions. For illustrative purposes we have
considered the simple bosonic QFT defined in (\ref{6}). However,
this model is interesting by itself since it is a bosonized
version of the Tomonaga-Luttinger model with a distance-dependent
forward-scattering potential. We obtained the central charge c as
function of $\sigma$. We predict $c=0$ for a very weakly decaying
potential corresponding to $\sigma< 1$. According to the results
of Schulz \cite{Schulz} this corresponds to the regime in which
the system behaves like a 1D version of a Wigner crystal. For
$\sigma\geq1$ we obtain $c=1$, in agreement with "common
knowledge" on Luttinger liquids \cite{Mironov-Zabrodin}. However,
in previous computations the value of c was inferred from the
scaling dimensions, whereas we are now providing a direct
determination for a whole family of well specified Coulombian
potentials. In contrast to all previous studies of CI in
long-ranged models we have explicitly computed correlators for
{\it different spatial and temporal points}, which allowed us to
capture the effect of anisotropy on c in a more complete fashion.
Moreover, by considering as a mathematical tool a deformation of
the critical theory given by (\ref{14}) and using a quantitative
prediction relating c with properties of the deformed model in the
scaling regime, we showed how, for $\sigma>1$, the correlation
length acquires a finite, interaction dependent renormalization
factor. If one naively ignores this factor, an unphysical,
interaction dependent conformal anomaly is obtained.

\vspace{1cm}

{\bf Acknowledgements}

The authors are grateful to Daniel Cabra, Eduardo Fradkin and
Gerardo Rossini for valuable discussions.

This work was partially supported by Universidad Nacional de La
Plata (Argentina) and Consejo Nacional de Investigaciones Cient\'
ificas y T\'ecnicas, CONICET (Argentina).


\begin{thebibliography}{99}

\bibitem{reviews} J. L. Cardy, in Phase Transitions and Critical
Phenomena, edited by C. Domb and J. L. Lebowitz (Academic, New
York, 1987), Vol. 11.\\
D. Boyanovsky and C. M. Na\'on, Rivista del Nuovo Cimento {\bf 13N2}, 1-76, (1990).\\
P. Di Francesco, P. Mathieu and D. S\'en\'echal, Conformal Field
Theory (Springer, New York, 1997).
\bibitem{Affleck-Cardy} H. W. J. Bl\"ote, J. L. Cardy and M. P. Nightingale, Phys. Rev. Lett. {\bf 56}, 7462 (1986).\\
I. Affleck, Phys. Rev. Lett. {\bf 56}, 746 (1986).
\bibitem{FQS} D. Friedan, Z. Qiu and S. Shenker, Phys. Rev. Lett. {\bf 52}, 1575
(1984).
\bibitem{Cardy3} J. L. Cardy, J. Phys. A {\bf 20}, L891 (1987).
\bibitem{Cardy4} J. L. Cardy, Phys. Rev. Lett. {\bf 60}, 2709 (1988).
\bibitem{Sutherland1} B. Sutherland, J. Math. Phys. {\bf 12}, 246
(1971); {\bf 12}, 251 (1971); Phys. Rev. A {\bf 4}, 2019 (1971);
{\bf 5}, 1372 (1972).
\bibitem{Haldane1} F. D. M. Haldane, Phys. Rev. Lett. {\bf 60}, 635
(1988).\\
B. S. Shastry, Phys. Rev. Lett. {\bf 60}, 639 (1988).
\bibitem{Haldane2} F. D. M. Haldane, Phys. Rev. Lett. {\bf 66},
1529 (1991).
\bibitem{Kawakami} N. Kawakami and Sung-Kil Yang, Phys. Rev. Lett. {\bf 67}, 2493
(1991).
\bibitem{Sutherland2} R. A. R\"omer and B. Sutherland, Phys. Rev. B,
{\bf 48}, 6058.
\bibitem{longrange}A.R. Goni, A. Pinczuk, J. S. Weiner, J. M. Calleja, B. S. Dennis, L. N. Pfeiffer, and K. W. West, Phys. Rev. Lett. {\bf67},
3298 (1991).\\ B. Dardel, D. Malterre, M. Grioni, P. Weibel, Y. Baer, J. Voit and D. Jerome, Europhys. Lett. {\bf 24}, 687
(1993); M. Nakamura, A. Sekiyama, H. Namatame, A. Fujimori, H. Yoshihara, T. Ohtani, A. Misu and M. Takano, Phys. Rev. B {\bf 49}, 16191(1994); A.
Sekiyama, A. Fujimori, S. Aonuma, H. Sawa, and R. Kato, Phys. Rev. B {\bf 51}, 13899 (1995).
\bibitem{spin-chains}J. Rodrigo Parreira, O. Bolina and J. Fernando Perez, J. Phys. A {\bf 30}, 1095, (1997).\\
Eddy Yusuf, Anuvrat Yoshi and Kun Yang, Phys. Rev. B {\bf 69}, 144412, (2004).
\bibitem{RKKY}M. A. Ruderman and C. Kittel, Phys. Rev. {\bf 96}, 99, (1954).\\
T. Kasuya, Progr. Theor. Phys. {\bf 16}, 45, (1956).\\
K. Yosida, Phys. Rev. {\bf 106}, 893, (1957).
\bibitem{Mironov-Zabrodin} A. D. Mironov and A. V. Zabrodin, Phys.
Rev. Lett. {\bf 66}, 534, (1991).
\bibitem{Li-Naon} K. Li and C. M. Na\'on, J. of Phys. A {\bf 31},
7929, (1998).
\bibitem{Schulz} H. J. Schulz, Phys. Rev. Lett. {\bf 71}, 1864,
(1993).
\bibitem{iucci-naon} A. Iucci and C. M. Na\'on, Phys. Rev. B {\bf 61}, 015530 (2000).
\bibitem{finite-temperature} T. Matsubara, Prog. Theor. Phys. {\bf 14}, 351, (1955). \\
C. Bernard, Phys. Rev. D {\bf 9}, 3312,
(1974).\\
 L. Dolan
and R. Jackiw, Phys. Rev. D {\bf 9}, 3320, (1974).
\bibitem{Manias} M. V. Man\'{\i}as, C. M. Na\'on and M. L. Trobo,
Nucl. Phys. B {\bf 525} [FS], 721, (1998).
\bibitem{Cardy5} J. L. Cardy, The Stress Tensor in Quenched Random
Systems, in Proceedings of the NATO Advanced Research Workshop on
Statistical Field Theories, Como 2001, (Kluwer, Boston, 2002).
\bibitem{Gurarie} V. Gurarie and A. W. W. Ludwig, hep-th/0409105.

\end{thebibliography}
\end{document}